\documentclass[review,5p,times,onecolumn,preprint]{elsarticle}
\usepackage{graphicx}
\usepackage{amssymb}
\usepackage{lineno}
\usepackage[version=3]{mhchem} 
\usepackage {xcolor}

\biboptions{numbers,sort&compress}

\journal{Surfaces and Interfaces}

\begin{document}

\begin{frontmatter}

\title{Polarized neutron reflectometry study from iron oxide nanoparticles monolayer}

\author[label1,label5]{V. Ukleev}
\ead{ukleev@lns.pnpi.spb.ru}
\address[label1]{B. P. Konstantinov Petersburg Nuclear Physics Institute, National Research Center "Kurchatov Institute", 188300 Gatchina, Russia}
\address[label5]{RIKEN Center for Emergent Matter Science (CEMS), Wako 351-0198, Japan}
\author[label3]{I. Snigireva}
\author[label3,label4]{A. Vorobiev}
\ead{avorobiev@ill.fr}
\address[label3]{European Synchrotron Radiation Facility, 71, Avenue des Martyrs, CS40220, F-38043 Grenoble Cedex 9, France}
\address[label4]{Department of Physics and Astronomy, Uppsala University, Box 516, 751 20, Uppsala, Sweden}

\begin{abstract}
We report on the polarized neutron reflectometry investigation of monolayer of magnetic iron oxide nanoparticles assembled by the Langmuir-Schaefer method. After deposition onto a solid substrate the polarized neutron reflectometry measurements in the external magnetic field were carried out. Thickness, density, roughness and in-depth resolved magnetization profile of the resulted layer were obtained from accurate fitting routine.  
\end{abstract}
\begin{keyword}
Magnetic nanoparticles \sep polarized neutron reflectometry \sep Langmuir film

\end{keyword}

\end{frontmatter}

\section{Introduction}
Arrays of ordered magnetic nanoparticles (MNPs) are promising candidates for a number of applications in the various fields including fundamental research \cite{gubin2005magnetic}, catalysis \cite{roy2009functionalized,rybczynski2003large,schatz2010nanoparticles}, sensors \cite{pankhurst2003applications,koh2009magnetic},  nanoelectronics and high-density data storage \cite{wen2011ultra,reiss2005magnetic,akbarzadeh2012magnetic,lu2007magnetic,nie2010properties,terris2005nanofabricated}. One of the straightforward ways to obtain ordered monolayer of MNPs is self-assembly of nanoparticles on a liquid subphase using Langmuir technique. In this routine, self-organization process occurs during isothermal compression of the Langmuir layer. However, interactions between MNPs has a non-trivial dependence on the particles size and shape, what results in different types of spacial ordering. 

Scanning electron microscopy is widely used to characterize a final structure of MNP monolayers after a deposition on solid substrates \cite{guo2003patterned,wen2011ultra}. To characterize structure of Langmuir monolayers both in-situ, i.e. on a liquid subphase, and ex-situ, i.e. on a solid substrate, X-ray scattering and reflectometry can be employed \cite{wang2012interfacial,stefaniu2014x,al2013situ,vorobiev2015substantial,ukleev2016x}. However, none of these methods provides information about magnetic arrangement, except state of the art experiments on electron holography \cite{yamamoto2008direct,yamamoto2011dipolar} and, possibly, resonant magnetic X-ray scattering (such experiments has not been reported so far according to our knowledge). Although, integrated magnetic signal from monolayer can be measured by high-sensitive SQUID magnetometry technique \cite{lee2007vast,pichon2011tunable,pauly2012size}, distribution of magnetic moments inside of MNP monolayer is under discussion \cite{disch2012quantitative}. In this context, polarized neutron reflectometry (PNR) is the only experimental technique which is sensitive to both structural and magnetic internal arrangement of MNPs in a monolayer. However, the \textit{polarized} neutron reflectometry measurements on a \textit{single} monolayer of MNPs remains challenging due to an extremely small volume of scattering material. Moreover, magnetic moments of MNPs can be significantly reduced, as compared to the bulk materials or thin films, due to the finite-size effects and surface (re-)oxidation, what especially concerns small MNPs. 

Previously, D. Mishra et al. reported assembly of iron oxide MNPs monolayer on silicon substrates and on vanadium films via spin-coating technique along with the further characterization by PNR \cite{mishra2015polarized}. However, in both cases it was not possible to avoid overlaying of a complete MNPs monolayer with less ordered patches of a second layer. Along with this one should note, that results of successful PNR measurements on \textit{multilayers} of relatively large MNPs (5 layers magnetite particles 20 nm in diameter) or on relatively thick films composed of MNPs (few hundreds nanometers of cobalt particles 13 nm in diameter) can be found in works \cite{mishra2012self} and \cite{theis2008self,theis2011self}  respectively.

In present study we investigated a monolayer of magnetic iron oxide MNPs 10 nm in diameter assembled on water surface and deposited onto a solid substrate by means of Langmuir-Schaefer technique. The structural and magnetic in-depth-resolved profiles of resulted monolayer have been obtained using PNR technique.

\section{Materials and methods}

Chloroform solution of iron oxide MNPs with mean diameter $d=10$ nm and size tolerance 2.5 nm were purchased from Ocean Nanotech. Due to re-oxidation, in such nanoparticles, originally synthesized as magnetite (Fe$_3$O$_4$), the inner magnetite core is usually surrounded by a maghemite (Fe$_2$O$_3$) shell \cite{carvalho2013iron,santoyo2011magnetic}. Both types of iron oxides are magnetic, although maghemite possesses lower magnetic moment comparing to magnetite. To prevent coagulation due to dipole-dipole interaction, the iron oxide particles are stabilized by a monolayer of oleic acid (C$_{18}$H$_{33}$COOH) with corresponding thickness of 2 nm.

After assembling a monolayer in a Langmuir through, it was deposited on a solid substrate of lateral size $20\times 20$ mm using Langmuir-Schaefer technique (stamping). As a substrate we used silicon coated with gold. A buffer titanium layer was introduced in-between to improve adhesion and homogeneity of gold layer. The top surface of gold was functionalized with a layer of 1-pentadecanethiol molecules making it hydrophobic for better adhesion of the MNPs. Nominal thicknesses of Ti and Au layers were 5 nm and 10 nm respectively. The details of monolayer assembling, choice of the substrate and deposition routine are discussed elsewhere \cite{vorobiev2015substantial,ukleev2016x}. 

\begin{figure}[ht] 
\includegraphics[width=8.5 cm]{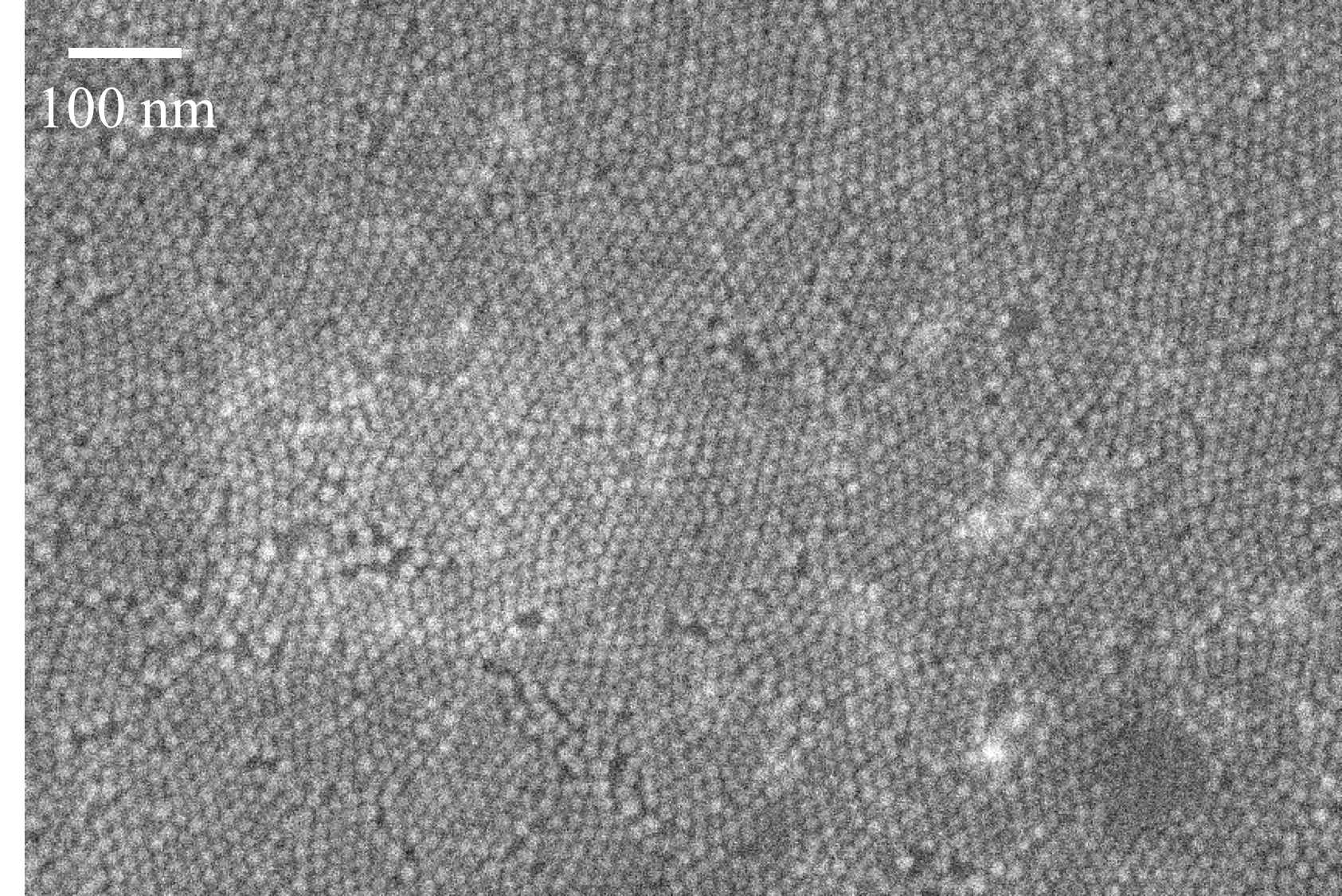}
\caption{SEM image of the monolayer transferred to the substrate.}
\label{sem} 
\end{figure}

PNR experiments were carried out on neutron reflectometer Super ADAM \cite{vorobiev2015recent} (Institut Laue-Langevin, Grenoble, France) using monochromatic neutron beam with wavelength $\lambda=5.18$ \AA ~and polarization $P_0=99.8~\%$. The detailed description of polarized reflectometry technique can be found elsewhere \cite{ankner1999polarized}. Reflectivity data sets were collected at room temperature in external magnetic field $B=500$ mT applied by means of an electromagnet either parallel ($R^+$ component) or anti-parallel ($R^-$ component) in respect to the vector  of incoming neutron polarization.

Modeling and fitting of the experimental data were performed using GenX package \cite{bjorck2007genx}. PNR datasets $R^+$ and $R^-$ were fitted simultaneously providing nuclear ($\rho_n$) and magnetic ($\rho_m$) components of scattering length density as a function of a distance $z$ from the surface of Si crystal. 

\begin{figure}[ht] 
\includegraphics[width=8.5 cm]{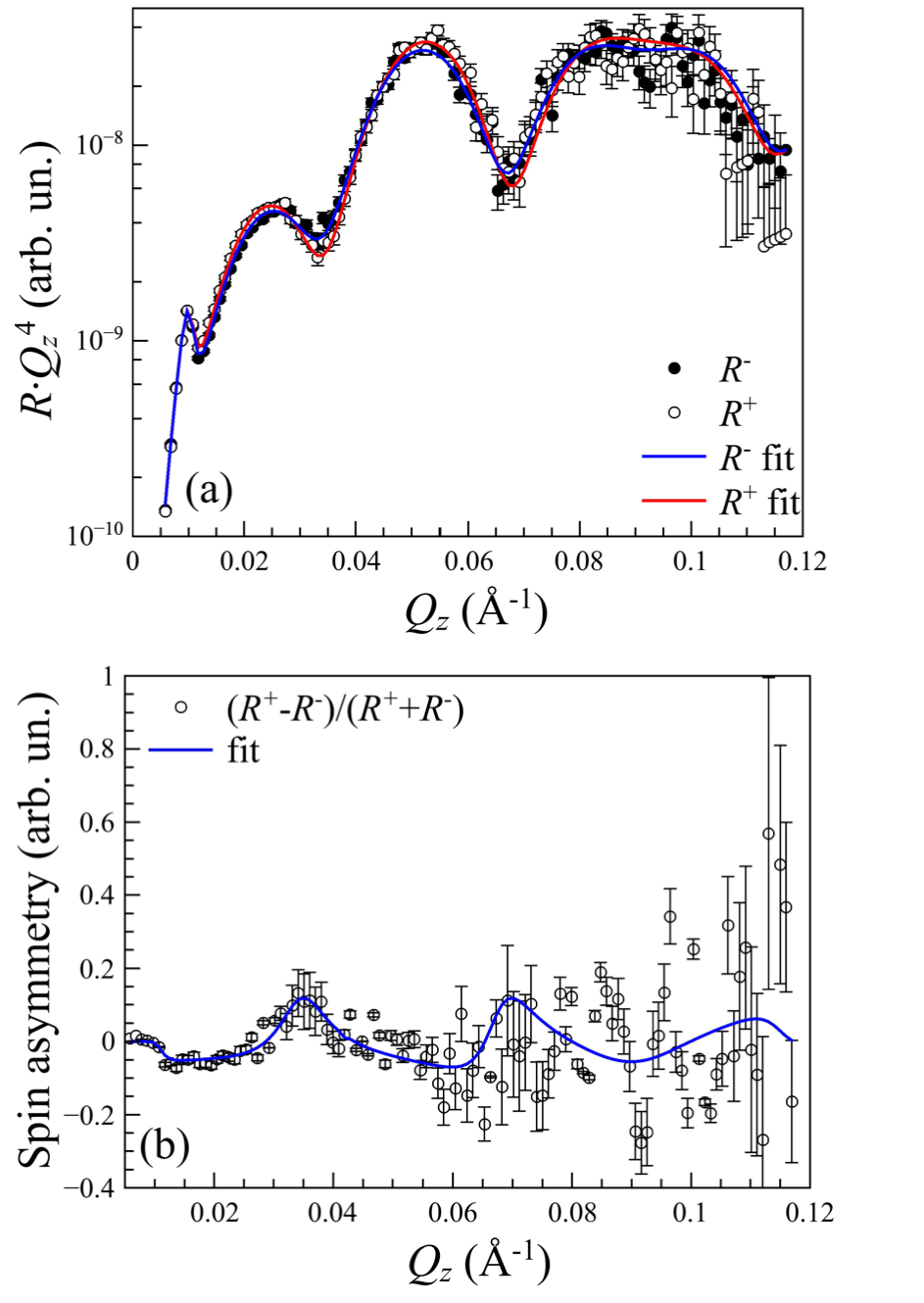}
\caption{(a) PNR curves and (b) spin asymmetry measured at applied $B=500$ mT. Symbols are corresponding to the measured data points, while the calculated curves are shown as the solid lines. PNR curves are multiplied by $Q_z^4$ to compensate a general decay of the reflectivity.}
\label{pnr}
\end{figure}

\begin{figure}[ht] 
\includegraphics[width=8.5 cm]{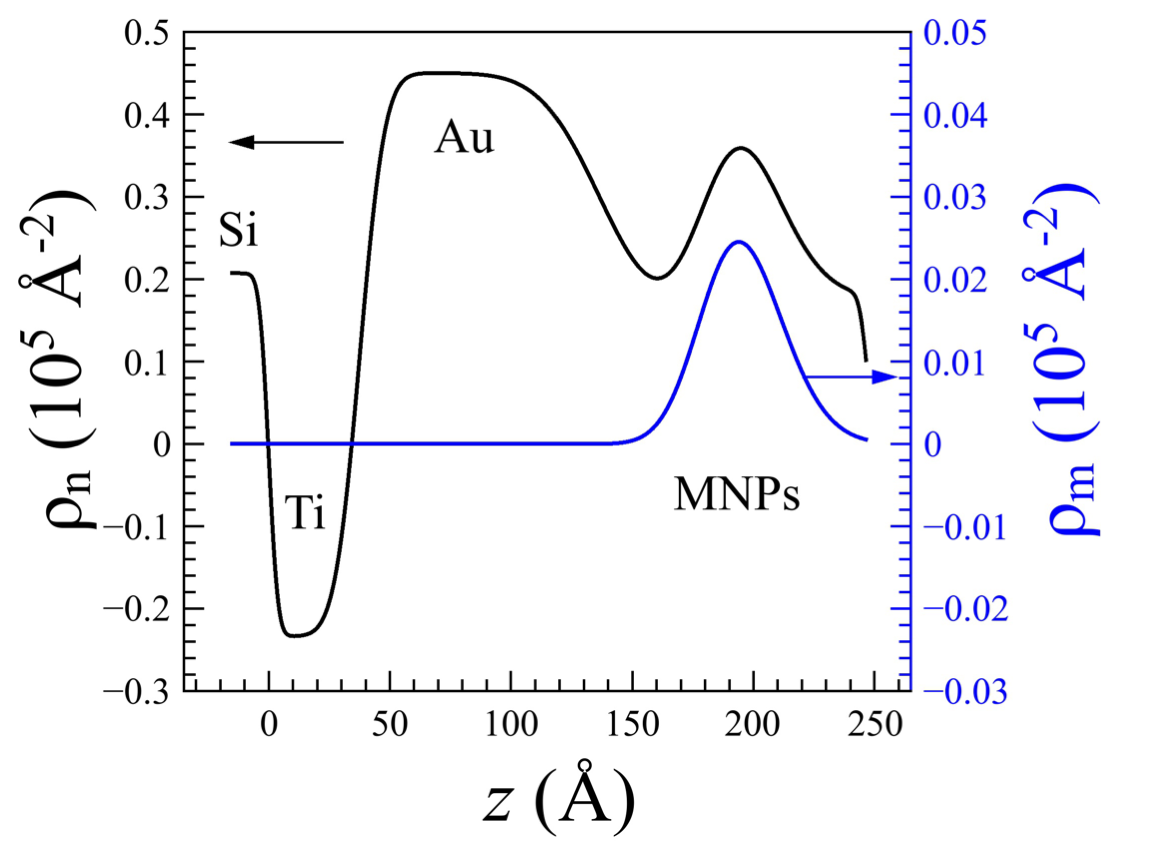}
\caption{Nuclear (black line) and magnetic (blue line) SLD profiles of the iron oxide monolayer obtained from the fitting routine. Zero point $z=0$ corresponds to the surface of Si crystal.}
\label{sld}
\end{figure}

\section{Results and discussion}

Measured and model PNR curves are shown in Fig.\ref{pnr}a as $R\cdot Q_z^4(Q_z)$ what is commonly used to emphasis quality of the fit. Here $Q_z=\frac{4\pi}{\lambda}\sin(\alpha)$ is a component of the wave vector transfer normal to the sample surface, $\alpha$ is angle of incidence. Splitting between the reflectivity curves with opposite polarizations hardly distinguishable in Fig. \ref{pnr}a is clearly visualized in form of so-called "spin asymmetry" $SA=\frac{R^+-R^-}{R^++R^-}$ on (Fig. \ref{pnr}b). Deviation of $SA(z)$ from zero is an unmistakable indication of the sample magnetization.  

We fitted experimental data with a model consisting of five layers on the top of Si crystal. Two of them described Ti and Au layers of the substrate, while three other aimed to describe the MNP layer. Ti and Au layers of the model were characterized by three parameters: thickness, roughness, and nuclear SLD -- $\rho_n$. Nuclear SLD values for Si, Ti and Au layers were fixed at corresponding table values \cite{varley1992neutron} while thicknesses and roughnesses were fitted. For layers describing magnetic nanoparticles both nuclear and magnetic SLDs ($\rho_n$ and $\rho_m$, respectively) were introduced. The latter one is directly proportional to the net magnetization $M$ of corresponding layer:
\begin{equation}
\rho_m=\frac{4 \pi m_n \mu_n M}{2 \pi \hbar^2},
\end{equation}
where $m_n$ is the neutron mass and $\mu_n$ is the magnetic moment of neutron. All four parameters for each of three layers describing the MNPs were fitted.

As the best fit $\rho_n(z)$ and $\rho_m(z)$ model profiles are shown in Fig. \ref{sld}. Calculated reflectivity curves  excellently reproduce the experimental data: chi-square parameter for goodness of fit resulted to $\chi^2=1.08$. Error bars did not exceed 4\% for nuclear SLD values, 5\% for thicknesses and 15\% for roughnesses. Moreover, nuclear SLD profile is found to be in a good agreement with corresponding data obtained in the previous X-ray reflectometry experiments \cite{ukleev2016x}. 

The nanoparticles monolayer is manifested by a local increase of the nuclear SLD (black curve in Fig. \ref{sld}a) in a region $150<z<250$ \AA. Almost parabolic shape of the SLD distribution in this region obtained directly from the fitting routine is in a good agreement with expected one-dimensional projection of a spherical particle with diameter of 10 nm. Deviation from perfect parabolic function can be explained by small misalignment of the centers of MNPs in respect to the substrate surface and also by a non-perfect monodispersity of the particles ensemble. Compared to the previous work by D. Mishra et al. that exploited spin-coating technique \cite{mishra2015polarized}, we succeeded to obtain dense monolayer of magnetic iron nanoparticles without formation of the incomplete layers on top.

Calculated spin asymmetry presented by red line in Fig. \ref{pnr}b match very well corresponding experimental data. The resulted net magnetization in the center of the monolayer, corresponding the maximum value of $\rho_m(z)$ is $M=0.31 \pm 0.03$ $\mu_B$/f.u., or $21 \pm 1$ emu/g which is significantly lower compared to the bulk magnetization of maghemite $M_s$=78 emu/g at room temperature 
\cite{buschow2003handbook}. However, it should be taken into account, that MNPs occupy only a certain fraction of the monolayer volume. In a previous study on x-ray grazing incidence diffraction from the same sample, it was obtained that MNPs are laterally organized in hexagonal polycrystal with a mean lattice constant $a$=12 nm. Given by mean particle diameter $d=$10 nm it is easy to calculate that MNPs in the middle part of the layer (at $z\approx$200 \AA) occupy only $\approx$30\% of its volume. Thus maximum value of magnetization could not exceed $0.3\cdot M_s=$23 emu/g. Since this value is very close to the experimentally obtained $21 \pm 1$ emu/g, one can conclude that MNPs in the monolayer consist of pure maghemite and they are fully magnetized. It also follows from a parabolic shape of $\rho_m(z)$ function, repeating shape of $\rho_m(z)$ distribution in the range $150<z<250$ \AA, what implies that MNPs are magnetized homogeneously. In other words, (magnetite core)-(maghemite shell) model is not confirmed for this type of the MNPs. The most possible reason for this is small size of the particles.

\section*{Conclusion}
We examined a monolayer of 10 nm magnetic iron oxide nanoparticles at the solid substrate by means of polarized neutron reflectometry. Even with such a small amount of scattering material it was possible to perform a very accurate fitting routine. As a result we have obtained both nuclear and magnetic scattering density profiles of the nanoparticles monolayer. The value of saturation magnetization of nanoparticles monolayer $M=21\pm1$ emu/g was determined. The distinct shape of magnetic moment distribution inside the monolayer reproduces the nuclear scattering length density profile shape giving evidence of homogeneous magnetization of the nanoparticles, which  in this case should consist of pure ferromagnetic iron oxide. These results will be used in further theoretical and micromagnetic investigations of nanoparticles self-assembly.

\section*{Acknowledgments}

Authors thank Institut Laue Langevin for provided beamtime at Super ADAM reflectometer. This work was supported by the Russian Foundation for Basic Research (grant 14-22-01113-ofi m).

\bibliographystyle{model1a-num-names}
\bibliography{biblio}

\end{document}